\documentclass[journal]{IEEEtran}
\usepackage[utf8]{inputenc}
\usepackage[T1]{fontenc}
\usepackage{amsmath}
\usepackage{tabularx}
\usepackage{amsfonts}
\usepackage{amssymb}
\usepackage[version=4]{mhchem}
\usepackage{stmaryrd}
\usepackage{graphicx}
\usepackage[export]{adjustbox}
\graphicspath{ {./images/} }
\usepackage{caption}
\usepackage{hyperref}
\hypersetup{colorlinks=true, linkcolor=blue, filecolor=magenta, urlcolor=cyan,}
\usepackage{multirow}
\usepackage{bbold}
\usepackage{hyperref}
\usepackage[dvipsnames]{xcolor}
\usepackage{booktabs} 
\usepackage{multicol}
\usepackage{subcaption}
\usepackage{graphicx}
\usepackage{caption}
\usepackage{amsthm}

\newtheoremstyle{boldhead}
  {} % Space above
  {} % Space below
  {\normalfont} % Body font - non-italic
  {} % Indent amount
  {\bfseries\itshape} % Theorem head font - bold and italic
  {.} % Punctuation after theorem head
  { } % Space after theorem head
  {\thmname{#1}\thmnumber{ #2}\thmnote{ \normalfont(#3)}} % Theorem head spec (for definitions and remarks)

\theoremstyle{bolditalichead}
\newtheorem{theorem}{Theorem}

\newtheorem{assumption}{Assumption}
\newtheorem{definition}{Definition}
\newtheorem{remark}{Remark}

\graphicspath{ {./Figs/} }
\usepackage{amsfonts}
\usepackage{amssymb}

\usepackage[noend]{algpseudocode}
\makeatletter
\def\BState{\State\hskip-\ALG@thistlm}
\makeatother
\usepackage{cite}
\usepackage{paralist}
\usepackage{color}

\definecolor{blue}{rgb}{0, 0.1, 0.7}
\usepackage{mathrsfs}
\usepackage{multirow}
\usepackage{graphicx}
\usepackage{subcaption}
\usepackage{caption}
\usepackage[font=small,justification=justified,singlelinecheck=false]{caption}

\title{Decentralized Attack-Resilient CLF-Based Control of Nonlinear DC Microgrids under FDI Attacks}

\author{Mohamadamin~Rajabinezhad,
        Muratkhan~Abdirash,
        Xiaofan~Cui,
        Shan~Zuo%
\thanks{M. Rajabinezhad and S. Zuo are with the Department of Electrical and Computer Engineering, University of Connecticut, Storrs, CT, USA (email: mohamadamin.rajabinezhad@uconn.edu, shan.zuo@uconn.edu). M. Abdirash and X. Cui are with the Department of Electrical and Computer Engineering, University of California, Los Angeles, CA, 90095 USA (email: mabdirash@ucla.edu, cuixf@seas.ucla.edu)}{\thanks{\textbf{Accepted for presentation at IEEE PES General Meeting 2026.
© IEEE. Personal use permitted. Final version will appear in IEEE Xplore.}}
}
}

\let\svthefootnote\thefootnote
\newcommand\blfootnotetext[1]{%
  \let\thefootnote\relax\footnote{#1}%
  \addtocounter{footnote}{-1}%
  \let\thefootnote\svthefootnote%
}

\let\svfootnotetext\footnotetext
\renewcommand\footnotetext[2][?]{%
  \if\relax#1\relax%
    \ifnum\value{footnote}=0\blfootnotetext{#2}\else\svfootnotetext{#2}\fi%
  \else%
    \if?#1\ifnum\value{footnote}=0\blfootnotetext{#2}\else\svfootnotetext{#2}\fi%
    \else\svfootnotetext[#1]{#2}\fi%
  \fi
}

\begin{document}
\maketitle
\captionsetup{singlelinecheck=false}

% \small
\begin{abstract}
The growing deployment of nonlinear, converter-interfaced distributed  energy resources (DERs) in DC microgrids demands decentralized controllers that remain stable and resilient under a wide range of cyber–physical attacks and disturbances. Traditional droop or linearized control methods lack resilience and scalability, especially when the system operates in its nonlinear region or faces diverse false-data-injection (FDI) attacks on control inputs. In this work, we develop a Decentralized Attack-Resilien Control Lyapunov Function (AR-CLF) based Quadratic Program (QP) control framework for nonlinear DC microgrids that ensures large-signal stability in a fully decentralized manner. Built upon the port-Hamiltonian representation, the proposed controller dynamically compensates diverse attacks including exponentially unbounded control-input perturbations beyond the bounded-attack regime commonly assumed in existing methods, through an adaptive resilience term, without requiring global information. Simulations validate that the AR-CLF based QP controller achieves superior stability and resilience against unbounded attacks, paving the way for scalable, attack-resilient, and physically consistent control of next-generation DC microgrids.
\end{abstract}
\vspace{-3mm}
\section{Introduction}
DC microgrids provide an efficient, modular platform for integrating renewable generation and electronic loads, but their converter-dominated and nonlinear nature creates stability challenges ~\cite{yuan2025large}. Constant-power loads, low inertia, fast switching dynamics, and cyber–physical attacks, such as false-data-injection (FDI) and denial-of-service (DoS) attacks on control input channels can introduce unbounded perturbations that severely degrade voltage regulation and current sharing. To improve resilience, recent work has explored adaptive, robust, and event-triggered controllers. Examples include resilient event-triggered control under DoS attacks~\cite{hong2025resilient}, and adaptive sampled-data security control for stochastic actuator failures~\cite{wang2025adaptive}. Nonlinear and decentralized stabilization methods based on Lyapunov optimization and port-Hamiltonian modeling have also been developed~\cite{chang2021potential,AbdirashACC}, along with decentralized secondary control under unreliable communication. Despite these developments, most existing designs fundamentally rely on the \emph{bounded-disturbance assumption}, limiting their effectiveness against control input-channel attacks whose magnitude increases over time.

CLF provide a rigorous foundation for guaranteeing stability in nonlinear systems and support both continuous and optimization-based feedback synthesis~\cite{barroso2023clf}. Although Input-to-state and robust CLF extensions can accommodate parametric uncertainty and bounded disturbances~\cite{zhao2025pdclf, nguyen2015clfqp}, they lose effectiveness under unbounded disturbances such as escalating FDI attacks. Consequently, existing CLF- and passivity-based controllers for DC microgrids guarantee stability only for bounded perturbations or local operating regions.
To overcome these limitations, this paper introduces a Decentralized AR-CLF based QP framework for nonlinear DC microgrids. The proposed method embeds resilience directly into the Lyapunov derivative through an adaptive compensation mechanism, ensuring stability under broad range of FDI attacks, including polynomially or exponentially unbounded attacks. Each converter employs a local AR-CLF based QP controller using only locally measurable variables, eliminating the need for centralized coordination and ensuring scalability across large-scale microgrids. 
The main contributions of this work are summarized as follows:
\begin{itemize}
    \item AR-CLF based QP Control: A fully decentralized AR-CLF based QP control framework is developed that ensures bounded convergence and stability under broad unbounded FDI attacks on control input channels, including exponentially growing attacks, without requiring centralized  communication or coordination.
    % \item \textbf{Decentralized Resilient CLF Control:} A fully decentralized attack-resilient CLF framework is proposed for nonlinear DC microgrids, ensuring stability under broad unbounded actuator disturbances, including exponentially growing FDI attacks.
    \item Nonlinear Port-Hamiltonian Modeling: The controller is formulated directly on the nonlinear averaged converter dynamics, preserving source–bus–load coupling and enabling global Lyapunov analysis without small-signal approximations.
    % \item \textbf{Attack-Resilient Convergence Guarantees:} Sufficient conditions are derived to ensure bounded convergence under unbounded FDI distortions, achieved without communication or centralized coordination.
   \item Large-Signal Stability with Rigorous CLF Analysis: A global energy-based CLF, derived from the port--Hamiltonian structure of the nonlinear DC microgrid, provides large-signal stability guarantees for all converter subsystems. A rigorous Lyapunov-based proof framework certifies stability of the proposed resilient controller, guaranteeing convergence even under unbounded FDI attacks to control input.
\end{itemize}
\vspace{-3mm}
\section{ Notions and Tools from Control Theory}
Consider the following control--affine system
\begin{equation}
\dot{x}=f(x)+g(x)u, 
\label{eq:affine}
\end{equation}
where its state and control input are given by 
$x \in X \subseteq \mathbb{R}^{n}$ and $u \in U \subseteq \mathbb{R}^{m}$, respectively. 
A pair $(u^{*}, x^{*}) \in U \times X$ is an equilibrium of system~\eqref{eq:affine}, if $f\left(x^{*}\right)+g\left(x^{*}\right) u^{*}=0_{n}$. 

% \begin{definition} Let the gradient operator $\partial(\cdot)$ be with respect to $x$ and in the column-vector form. A Lie derivative is an operator acting on a function $V$ and is defined as $\mathcal{L}_{f} V=\partial V(x)^{\top} f(x)$.
% \end{definition}

% \begin{definition} Let $V: X \rightarrow \mathbb{R}$ be a real-valued function. $V$ is Lipschitz continuous on the domain $X$, if there exists a constant $M>0$ such that $\|V(x)-V(y)\| \leq M\|x-y\|, \forall x, y \in X$, where $\|\cdot\|$ is the Euclidean norm.
% \end{definition}

% \begin{definition} Let $k(x)$ be a Lipschitz continuous feedback control law for (1) with $k\left(x^{*}\right)=u^{*}$. The equilibrium $x^{*}$ of the closed-loop system $\dot{x}=f(x)+g(x) k(x)$, is (locally) exponentially stable, if there exist constants $M, \lambda>0$ such that $ \left\|x(t)-x^{*}\right\| \leq M\left\|x_{0}-x^{*}\right\| e^{-\lambda t}$, holds $\forall t>0$ and $\forall x_{0} \in D \subset X$ where $x(t)$ is the solution of \eqref{eq:affine} to an initial condition $x_{0}$.
% \end{definition}
%%%Shorter version of def 1
% \begin{definition}
% Let $k(x)$ be a Lipschitz feedback law with $k(x^*)=u^*$.  
% The equilibrium $x^*$ of $\dot{x}=f(x)+g(x)k(x)$ is (locally) exponentially stable if there exist $M,\lambda>0$ such that $\|x(t)-x^*\|\le M\|x_0-x^*\|e^{-\lambda t}, \quad \forall t>0,\ \forall x_0\in D\subset X$.
% \end{definition}
\begin{definition}\label{def:CLF} \cite{ames2016control}
A continuously differentiable function
$V:\mathbb{R}^{n}\to\mathbb{R}_{\ge 0}$ is called a
\emph{Control Lyapunov Function}
if there exist positive constants $c_1,c_2,c_3>0$ such that
\begin{equation}
c_1\|x\|^2 \;\le\; V(x) \;\le\; c_2\|x\|^2, \quad \forall x\in\mathbb{R}^n,
\end{equation}
\begin{equation}
\inf_{u\in\mathbb{R}^m}
\bigl[L_f V(x) + L_g V(x)\,u + c_3 V(x)\bigr] \;\le\; 0,
\quad \forall x \neq 0.
\end{equation}
Here, $L_f V$ and $L_g V$ denote the Lie derivatives of $V$ along $f$ and $g$, respectively.
\end{definition}
\begin{definition}\label{Def_Port-Hamiltonian}\cite{ortega2002interconnection}
A Port-Hamiltonian (PH) dynamical system is given by
\begin{align}
& \dot{x}=(\mathcal{J}(x)-\mathcal{R}(x)) \partial H(x)+g(x) u+g_{z}(x) z,  \tag{3a}\\
& y=g_{z}(x)^{T} \partial H(x), \tag{3b}
\end{align}
where skew-symmetric $\mathcal{J}(x)^{T}=-\mathcal{J}(x)$ is called an interconnection matrix, and positive definite $\mathcal{R}(x)^{T}=\mathcal{R}(x) \geq 0$ is called a dissipation matrix. The function $H$ is called a Hamiltonian and is defined as $H(x)=x^{T} Q x$ with positive definite $Q^{T}=Q>0$. Let $z \in Z \subseteq \mathbb{R}^{p}$ denote the input that enters the system modulated by matrix $g_{z}(x)$, and $y \in Y \subseteq \mathbb{R}^{p}$ denote the output.
\end{definition}
\begin{definition} \label{Def_GIPH} \cite{AbdirashACC} Consider $k$ interconnected PH subsystems (Def. \ref{Def_Port-Hamiltonian} indexed by $j=1, \ldots, k$) such that all of their input and output ports become internal, i.e. $\sum_{j=1}^{k} y_{j}^{T} z_{j}=0$. Then the Global Interconnected Port-Hamiltonian System is given by 
\vspace{-1em}
\begin{align} &x =\operatorname{col}\left(x_{1} \ldots x_{k}\right), H(x)  =\textstyle \sum_{j=1}^{k} H_{j}\!\left(x_{j}\right)\tag{4a}\\
&\mathcal{R}(x)  =\operatorname{diag}\left(\mathcal{R}_1\left(x_{1}\right) \ldots \mathcal{R}_{k}\left(x_{k}\right)\right), \tag{4b} 
\end{align}
where $\operatorname{col}(\cdot)$ stacks the column vectors and $\operatorname{diag}(\cdot)$ constructs a block-diagonal matrix. The global $u$ is obtained by stacking $u_i$, while the global $\mathcal{J}(x)$ and $g(x)$ are both block-diagonal with entries $\mathcal{J}_j(x)$ and $g_j(x)$, respectively.  
\end{definition}
\begin{definition}[\cite{khalil2002nonlinear}]
\label{def:3}
$x(t) \in \mathbb{R}^{n}$ is  uniformly ultimately
bounded (UUB) with the ultimate bound $b$, if there exist constants $b, c>0$, independent of $t_{0} \geq 0$, and for every $a \in(0, c)$, there exists $t_{1}=t_{1}(a, b) \geq 0$, independent of $t_{0}$, such that
% \begin{equation}
% \label{UUB_eq}
$\;\|x\left(t_{0}\right)\| \leq a \Rightarrow\|x(t)\| \leq b, \; \forall t \geq t_{0}+t_{1}$.
% \end{equation}    
\end{definition}
\section{Preliminaries on DC Microgrid Modeling and Optimal Steady-State Analysis}
\begin{figure}[t]
    \centering
    \includegraphics[scale=0.52]{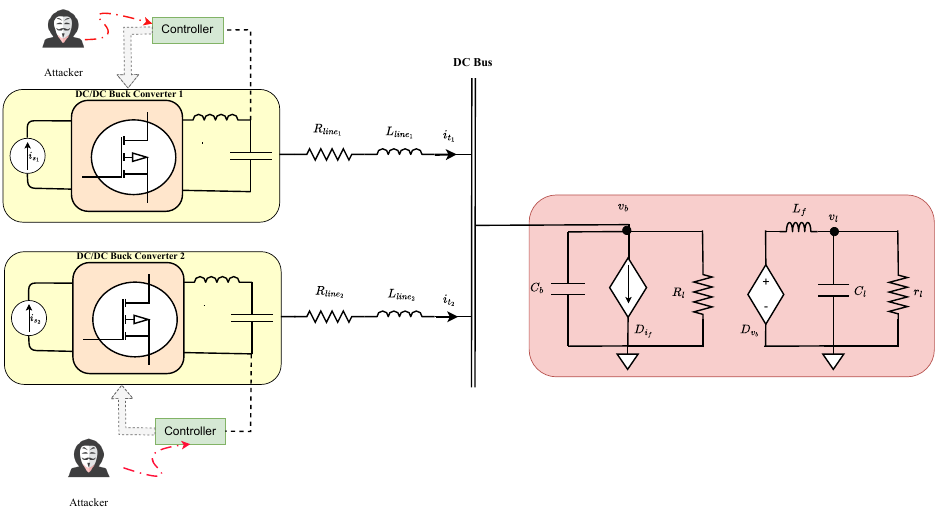}
    \caption{Circuit schematics of a single-bus DC microgrid.}
    \label{DC MG_structure}
\end{figure}
This section models a DC microgrid that operates in islanded mode and consists of $(k-1)$ decentralized  energy resources (DERs), where $k \ge 1$. DERs interface with a common DC bus through DC/DC converters. By averaging the switched-circuit dynamics of each converter, a reduced continuous-time model is obtained. On the source side, the combined DER plus converter dynamics are represented by a controlled current source $i_{s_j}$ in parallel with an output capacitor $C_j$ (see Fig.~\ref{DC MG_structure}). Under current-mode operation, the fast inner loop enforces the inductor current reference, while the outer loop provides this reference value; hence, $i_{s_j}$ is treated as an instantaneous control input in the averaged model. The load components are aggregated into linear and nonlinear parts. The linear portion is modeled by a resistor $R_l$, whereas the nonlinear portion employs a DC/DC converter that actively adjusts its duty ratio to regulate voltage. This converter is equivalently represented by a DC transformer with a turn ratio $0 \le d_l \le 1$, followed by a low-pass filter composed of $L_f$, $C_l$, and a load $r_l$, whose states are the inductor current $i_f$ and the output voltage $v_l$.

For decentralized analysis and control, the microgrid is partitioned into $k$ dynamically interconnected subsystems. Each source subsystem ($j = 1, \ldots, k-1$) includes the converter and its associated line. The control input is $i_{s_j}$, the coupling signal is the common bus voltage $v_b$, and the subsystem output is the line current $i_{t_j}$. By applying Kirchhoff’s voltage and current laws, the dynamics of subsystem $j$ are described by
% \vspace{-0.2em}
\begin{equation}
\label{eq:source_subsystem}
\begin{bmatrix}
C_j\dot{v}_j \\[2pt]
L_j\dot{i}_{t_j}
\end{bmatrix}
=
\begin{bmatrix}
- i_{t_j}\\[2pt]
v_j - R_j i_{t_j} - v_b
\end{bmatrix}
+
\begin{bmatrix}
1 \\[2pt]
0
\end{bmatrix}
i_{s_j},
\end{equation}
where all subsystems are coupled through $v_b$. The final subsystem ($j = k$) represents the bus and load converter dynamics. Its control variable is the duty cycle $d_l$, while the external interaction arises from $\sum_{j=1}^{k-1} i_{t_j}$. Applying Kirchhoff’s laws yields
\vspace{-0.5em}
\begin{equation}
\label{eq:load_subsystem}
\begin{bmatrix}
C_b\dot{v}_b \\[2pt]
L_f\dot{i}_f \\[2pt]
C_l\dot{v}_l
\end{bmatrix}
=
\begin{bmatrix}
\sum_{j=1}^{k-1} i_{t_j} - {v_b}/{R_l}\\[2pt]
- v_l \\[2pt]
{i_f - v_l / r_l}
\end{bmatrix}
+
\begin{bmatrix}
- {i_f} \\[2pt]
{v_b} \\[2pt]
0
\end{bmatrix}
d_l.
\end{equation}

%\section{Optimal Steady-State Behavior}
The equilibrium points of the DC microgrid can be determined by solving the steady-state conditions of \eqref{eq:source_subsystem} and \eqref{eq:load_subsystem}. However, this results in an underdetermined set of equations, yielding infinitely many equilibria parameterized by the steady-state current distribution, the bus voltage $v_b^{*}$, and the duty ratio $d_l^{*}$ of the load converter. To obtain a unique and optimal operating point, a strictly convex optimal power flow (OPF) problem is formulated to minimize total steady-state losses in the network. These losses are primarily caused by power dissipation across the resistances $R_j$ of the transmission lines. The steady-state bus dynamics serve as the constraint, leading to the following optimization problem:
\vspace{-0.5em}
\begin{equation}
\label{eq:opf_problem}
\begin{aligned}
\min_{\{i_{t_j}^{*}\}_{j=1}^{k-1}}\sum_{j=1}^{k-1} (i_{t_j}^{*})^{2} R_j, 
\text{ subject to }\sum_{j=1}^{k-1} i_{t_j}^{*} =\tfrac{v_b^{*}}{R_l} + \tfrac{d_l^{*}v_b^{*}}{r_l}.
\end{aligned}
\end{equation}
The optimal solution of the OPF problem \eqref{eq:opf_problem} is obtained as
\begin{equation}
\label{eq:optimal_current}
i_{t_j}^{*} = \left({\tfrac{v_b^{*}}{R_l} + \tfrac{d_l^{*}v_b^{*}}{r_l}}\right)/\left({\textstyle \sum_{p=1}^{k-1}} \! \frac{R_j}{R_p}\right),
\end{equation}
where $j=1,\ldots,k$. The resulting steady-state distribution in \eqref{eq:optimal_current} ensures equitable current sharing among DERs in proportion to their line conductance, compelling each source to supply sufficient power to offset transmission losses. Consequently, all DER units contribute equally to the total load demand. The remaining steady-state quantities follow from the circuit relationships as
\vspace{-0.5em}
\begin{equation}
\label{eq:ss_values}
\begin{gathered}
v_j^{*} = i_{t_j}^{*} R_j + v_b^{*},
i_{s_j}^{*} = i_{t_j}^{*},
v_l^{*} = d_l^{*} v_b^{*},
i_f^{*} = {v_l^{*}}/{r_l}.
\end{gathered}
\end{equation}
\section{Decentralized Large-Signal Stable Controller Design}
\subsection{Port-Hamiltonian Representation of DC Microgrid}
A port-Hamiltonian (PH) system (see Definition ~\ref{Def_Port-Hamiltonian}) naturally admits a Lyapunov function derived from its Hamiltonian, which guarantees large-signal stability. Moreover, PH systems can be seamlessly interconnected via their input--output ports, making them well-suited for decentralized control design. The source and load dynamics in \eqref{eq:source_subsystem} and \eqref{eq:load_subsystem} can therefore be represented as coupled PH subsystems that collectively form a global interconnected PH model described in Definition \ref{Def_GIPH}. Consistent with Section~III, the DC microgrid consists of $k$ PH subsystems. For $j = 1, \ldots, k-1$, each subsystem represents a DER and its grid-interfacing converter, equivalent to~\eqref{eq:source_subsystem}, and is given by
\vspace{-0.5em}
\begin{subequations}
\label{eq:ph_source}
\begin{align}
&x_j = [v_j,\, i_{t_j}]^{\top}, u_j = i_{s_j}, z_j = -v_b, y_j = i_{t_j}, \label{eq:ph_source_a}\\
&H_j(x_j) = \tfrac{1}{2}\left(C_j v_j^2 + L_j i_{t_j}^2\right), \label{eq:ph_source_b}\\
%\partial H_j(x_j) &= [C_j v_j,\, L_j i_{t_j}]^{\top}, \label{eq:ph_source_c}\\
&\mathcal{J}_j= 
\begin{bmatrix}
0 & -1/L_j C_j\\[2pt]
1/L_j C_j & 0
\end{bmatrix},g_j = [1/C_j,\, 0]^{\top} \label{eq:ph_source_d}\\
&\mathcal{R}_j= \text{diag}(0,R_j/L_j^2),g_{z_j} = [0,\, 1/L_j]^{\top}.
%\begin{bmatrix}
%0 & 0\\[2pt]
%0 & R_j/L_j^2
%\end{bmatrix}, 
\label{eq:ph_source_e}
\end{align}
\end{subequations}
The load subsystem $k$, corresponding to \eqref{eq:load_subsystem}, captures the DC bus and load dynamics and is defined as
\begin{subequations}
\label{eq:ph_load}
\begin{align}
&x_k = [v_b,\, i_f,\, v_l]^{\top}, u_k = d_l, z_k = -[i_{t_1},\, \ldots,\, i_{t_{k-1}}]^{\top},\label{eq:ph_load_a}\\
&H_k(x_k) = \tfrac{1}{2}\left(C_b v_b^2 + L_f i_f^2 + C_l v_l^2\right), \label{eq:ph_load_b}\\
%\partial H_k(x_k) &= [C_b v_b,\, L_f i_f,\, C_f v_l]^{\top}, \label{eq:ph_load_c}\\
&\mathcal{J}_k=
\begin{bmatrix}
0 & 0 & 0\\
0 & 0 & -1/L_f C_l\\
0 & 1/L_f C_l & 0
\end{bmatrix}, \label{eq:ph_load_d}\\
&\mathcal{R}_k=\text{diag}\left(1/R_l C_b^2, 0, 1/r_l C_l^2\right),
%\begin{bmatrix}
%1/(R_l C_b^2) & 0 & 0\\
%0 & 0 & 0\\
%0 & 0 & 1/(r_l C_f^2)
%\end{bmatrix}, 
\label{eq:ph_load_e}\\
&g_k(x_k) = [-i_f/C_l,\, v_b/L_f,\, 0]^{\top}, \label{eq:ph_load_f}\\
&g_{z_k} = -[1/C_b,\, 0,\, 0]^{\top},y_k = -[v_b,\, \ldots,\, v_b]^{\top} \in \mathbb{R}_{\le 0}^{k-1}. \label{eq:ph_load_i}
\end{align}
\end{subequations}

The PH subsystems are interconnected through their ports. As $\sum_{j=1}^{k} y_j^{\top} z_j
= \sum_{j=1}^{k-1}(-i_{t_j} v_b + v_b i_{t_j}) = 0$, no external port remains, confirming that the complete microgrid forms a global interconnected PH system.
\vspace{-1em}
\subsection{Large-Signal Stable DC Microgrid}
Let $v_j^{*}$ and $i_{t_j}^{*}$ denote the desired steady-state voltage and current, respectively, obtained from~\eqref{eq:opf_problem} and~\eqref{eq:optimal_current}. To ensure exponential convergence to $x_j^{*}$, each source subsystem is shaped into a closed-loop PH structure with additional damping on the voltage term. Define the error coordinates as $\hat{x}_j = x_j - x_j^{*},\hat{u}_j = u_j - u_j^{*},$ and $ \hat{z}_j = z_j - z_j^{*}, \hat{y}_j = y_j - y_j^{*}$. Modified closed-loop matrices are then given by
\begin{subequations}
\label{eq:ph_source_cl}
\begin{align}
%\hat{x}_j &= x_j - x_j^{*}, \label{eq:ph_source_cl_a}\\[-2pt]
\mathcal{J}_j^{*} &=
\begin{bmatrix}
0 & -1/L_j C_j\\
1/L_j C_j & 0
\end{bmatrix}, \label{eq:ph_source_cl_b}\\[-2pt]
\mathcal{R}_j^{*} &= \text{diag}\left(\alpha_j/C_j^2,R_j/L_j^2\right),
%\begin{bmatrix}
%\alpha_j/C_j^2 & 0\\
%0 & R_j/L_j^2
%\end{bmatrix}, 
\label{eq:ph_source_cl_c}
%\hat{u}_j &= u_j - u_j^{*}, \quad
%\hat{z}_j = z_j - z_j^{*}, \quad
%\hat{y}_j = y_j - y_j^{*}. \label{eq:ph_source_cl_d}
\end{align}
\end{subequations}
Here, $\alpha_j > 0$ introduces voltage damping, ensuring exponential stability. For the load subsystem, similar shaping is applied by adding dissipation to the filter current and modifying the interconnection matrix. The closed-loop representation becomes
\vspace{-1em}
\begin{subequations}
\label{eq:ph_load_cl}
\begin{align}
%\hat{x}_k &= x_k - x_k^{*}, \hat{d}_l &= d_l - d_l^{*}, \label{eq:ph_load_cl_a}\\[-2pt]
\mathcal{J}_k^{*} &=
\begin{bmatrix}
0 & d_l^{*}/C_b L_f & 0\\
-d_l^{*}/C_b L_f & 0 & -1/C_l L_f\\
0 & 1/C_l L_f & 0
\end{bmatrix}, \label{eq:ph_load_cl_b}\\[-2pt]
\mathcal{R}_k^{*} &=\text{diag}\left(1/R_l C_b^2,\alpha_k/L_f^2, 1/r_l C_l^2\right).
%\begin{bmatrix}
%1/(R_l C_b^2) & 0 & 0\\
%0 & \alpha_k/L_f^2 & 0\\
%0 & 0 & 1/(r_l C_f^2)
%\end{bmatrix}, 
\label{eq:ph_load_cl_c}
%\hat{d}_l &= d_l - d_l^{*}. \label{eq:ph_load_cl_d}
\end{align}
\end{subequations}
\subsection{Nominal Stabilizing Controller}
The nominal controller follows the Dynamic Interconnection and Damping Assignment Passivity-Based Control (IDA-PBC) framework~\cite{yuan2025large}. Unlike the traditional approach, which may violate port passivity and lose Lipschitz continuity near equilibrium, the dynamic IDA-PBC formulation preserves both. It jointly solves for the control input $u_j$ and a transient auxiliary state $\tilde{x}_j$ such that the error becomes $\hat{x}_j=x_j-\tilde{x}_j-x^*_j$. %\begin{equation}
%\label{eq:IDA_condition}
%[\mathcal{J}_j^{*} - \mathcal{R}_j^{*}] \partial H_j(\hat{x}_j)
%- g_{z_j} z_j^{*} + \dot{\tilde{x}}_j
%= [\mathcal{J}_j - \mathcal{R}_j] \partial H_j(x_j)
%+ g_j u_j,
%\end{equation}
The resulting nominal stabilizing control laws are
\begin{align}
\label{eq:source_control}
&\hat{u}_j(\hat{x}_j) = -\alpha_j \hat{v}_j
+ e^{-R_j t / L_j} \hat{i}_{t_j}(0),\\
&\hat{d}_l(\hat{x}_k) =
\begin{cases}
-{\alpha_k/\hat{i}_f}{v_b}, & v_b > 0,\\
0, & v_b = 0.
\end{cases}
\end{align}
\section{Decentralized AR-CLF based Controller Design}
The nominal stabilizing controller presented in Section~IV guarantees convergence under normal operating conditions but does not provide resilience against malicious cyber attacks on control inputs. In practical DC microgrids, FDI attacks may corrupt locally transmitted control or state information. Since individual agents typically operate using only local measurements and limited neighborhood information, without access to global network topology or centralized supervision, such attacks cannot be reliably detected or mitigated through global monitoring or coordination. This motivates the development of a fully decentralized attack-resilient controller that embeds resilience guarantees directly within an AR-CLF framework.
\vspace{-1.5em}
\subsection{Stability via CLF}
Let the global Hamiltonian $H$ denote a candidate Lyapunov function certifying exponential stability of the desired equilibrium $x^{*}$ for the closed-loop DC microgrid. Following the same reasoning as in Section~IV, let $H(\hat{x}) = \tfrac{1}{2}\hat{x}^{\top} Q\,\hat{x}$, then it is easy to show that the global Hamiltonian is positive-definite and continuously differentiable, where $Q= \operatorname{diag}(C_{1}, L_{1}, \ldots, C_{k-1}, L_{k-1}, C_{b}, L_{f}, C_{l})$. Under the nominal stabilizing controller, derivative of the Hamiltonian satisfies 
\vspace{-1em}
\begin{equation}\label{eq:Hdot_global}
\dot{H}(\hat{x}) = {\textstyle \sum_{j=1}^{k}} \dot{H}_{j}\!\left(\hat{x}_{j}\right)
 = -\,\partial H(\hat{x})^{\top}\mathcal{R}^{*}\,\partial H(\hat{x}) < 0,
\end{equation}
establishing global exponential stability of $x^{*}$ in the absence of adversarial disturbances. Hence, $H(\hat{x})$ serves as a valid CLF for the global dynamics, satisfying
\begin{equation}\label{eq:clf_nominal}
\mathcal{L}_{f}H(\hat{x}) + \mathcal{L}_{g}H(\hat{x})\,u \le -\,\hat{x}^{\top} Q\,\hat{\mathcal{R}}^{*} Q\,\hat{x},
\end{equation}
where $\hat{\mathcal{R}}^{*} = \mathcal{R}^{*} - \Lambda > 0$ and $\Lambda$ is a positive diagonal tuning matrix.
\vspace{-1.5em}
\subsection{Resiliency via AR-CLF based QP}
While the nominal controller ensures stability in the ideal case, an adversarial FDI attack can inject an additive perturbation $\delta_{i}(t)$ into each control channel, where $\delta_{i}(t)$ is unknown, time-varying, and possibly unbounded.
\vspace{-0.5em}
\begin{equation}\label{eq:affine_under_attack}
\dot{x}_{i} = f_{i}(x_{i}) + g_{i}(x_{i})\big(u_{i} + \delta_{i}(t)\big),
\end{equation}
% An attack is considered bounded if $\|\delta_i(t)\| < \bar\eta$, for some $\bar\eta > 0$; otherwise, it is unbounded. The attackers aim to disrupt the cooperative regulation system by introducing these unbounded disturbances.

\begin{assumption}
\label{ass: attacks}
$\delta_{j}(t)$ is an exponentially unbounded signal. That is, its norms grow at most exponentially with time. For the convenience of mathematical stability analysis, we assume 
$\|\delta_{j} \|\le \gamma_{j} \exp(\kappa_{j}t) $, where $\gamma_{j}$ and $\kappa_{j}$ are positive constants.
\end{assumption}
This exponential envelope is introduced as a conservative worst-case upper bound to enable rigorous Lyapunov-based stability analysis, rather than as an exact model of adversarial behavior, and it upper-bounds a broad class of realistic attack growth patterns.
% This exponential envelope is introduced as a conservative analytical abstraction to enable rigorous Lyapunov-based analysis and should be interpreted as a worst-case upper bound rather than a precise model of adversarial behavior.
To maintain boundedness under such conditions, the control input is obtained by solving QP with an AR-CLF constraint:
\begin{equation}
\label{eq:resilient_qp}
\begin{aligned}
\min_{u_{j}\ge0}\;& (u_{j}-u_{\mathrm{nom},j})^{2} \\[3pt]
\text{s.t.}\;& 
L_{f_{j}}V_{j}(x_{j}) + L_{g_{j}}V_{j}(x_{j})u_{j}  \\
&\quad
+\,\frac{(L_{g_{j}}V_{j})(L_{g_{j}}V_{j})^{\!\top}}
        {\|L_{g_{j}}V_{j}\| + e^{-\alpha t}}\,e^{\rho_{j}(t)}
\;\le\; -\beta_{j}V_{j}(x_{j}).
\end{aligned}
\end{equation}
where $V_{j}(x_{j})$ is the local CLF derived from the subsystem Hamiltonian, $\beta_{j}>0$, $\alpha>0$, $q_j>0$, and $\rho_{j}(t)$ is an adaptive gain evolving as
\begin{equation}\label{eq:rho_dynamics}
\dot{\rho}_{j}(t) = q_j \|L_{g_{j}}V_{j}(x_{j})\|, \qquad \rho_{j}(0) = \rho_{j0} \ge 0.
\end{equation}

% The adaptive law~\eqref{eq:rho_dynamics} autonomously amplifies the stabilizing term in~\eqref{eq:resilient_constraint} when the effect of $\delta_{j}(t)$ increases, enhancing robustness without requiring explicit attack estimation. The QP~\eqref{eq:resilient_qp} minimizes the deviation from the nominal PH-based control $u_{\mathrm{nom},j}$ while preserving Lyapunov decay under corrupted inputs. Since all quantities depend solely on local measurements, the proposed DARC operates in a fully decentralized manner, enabling each converter to autonomously counteract adversarial perturbations and maintain resilient convergence under FDI attacks.

\begin{assumption}
\label{ass:nonvanishing_clf_gradient}
There exist constants $c_j>0$ and $\varepsilon_j>0$ such that 
$\|L_{g_j}V_j(x_j)\|\ge c_j$ for all $\|x_j\|\ge \varepsilon_j$.
\end{assumption}

% \begin{remark}
% This regularity condition guarantees that the control direction $g_j(x_j)$ 
% can effectively influence the CLF when the state is away from the equilibrium. 
% It ensures that the resilient term in the CLF--QP remains uniformly effective 
% outside the small neighborhood $\|x_j\|\le\varepsilon_j$.
% \end{remark}
\vspace{-0.5em}
% {\color{red}
% % Modified remark 1 as follows
% \begin{remark}
% This regularity condition ensures that, whenever the state is sufficiently away 
% from the equilibrium, the control input retains non-vanishing authority over the 
% CLF. In the context of DC/DC converter dynamics, this corresponds to the ability 
% of the duty ratio or current reference to inject or dissipate electrical energy 
% when voltage or current deviations are nonzero. As a result, the adaptive resilient 
% term in the AR-CLF based QP remains effective outside the neighborhood 
% $\|x_j\|\le\varepsilon_j$, enabling stabilization under adversarial attacks.
% \end{remark}
% }
% {\color{red}The AR-CLF based QP is low-dimensional, locally solvable from local measurements, scale-independent, and compatible with secondary control time scales.}
\begin{remark}
This regularity condition ensures that, when the state is sufficiently away from the equilibrium, the control input retains non-vanishing authority over the CLF. For DC/DC converters, this corresponds to the ability of the duty ratio or current reference to inject or dissipate energy under nonzero voltage or current deviations. Consequently, the adaptive resilient term in the AR--CLF--based QP remains effective outside the neighborhood $\|x_j\|\le\varepsilon_j$. Moreover, the resulting QP is low-dimensional, locally solvable using only local measurements, independent of the network size, and compatible with secondary control time scales.
\end{remark}

\begin{theorem}
Consider subsystem $j$ under FDI attacks to control input channels modelled in \eqref{eq:affine_under_attack}, 
and let $V_{j}(x_{j})$ be a positive-definite, radially unbounded CLF. 
Under Assumption~\ref{ass: attacks} and the non-vanishing CLF gradient condition of 
Assumption~\ref{ass:nonvanishing_clf_gradient}, the decentralized AR-CLF based QP controller 
\eqref{eq:resilient_qp}, equipped with the adaptive gain dynamics \eqref{eq:rho_dynamics}, 
guarantees that subsystem~$j$ is UUB. In particular, the trajectories converge to an invariant set contained in the compact region 
$\|x_{j}\|\le\varepsilon_{j}$ despite broad range of FDI attacks including exponentially growing FDI signals. 
Furthermore, since \eqref{eq:resilient_qp} minimizes the deviation from the nominal PH-based 
control input $u_{\mathrm{nom},j}$ while maintaining Lyapunov decay, the controller preserves 
nominal performance under normal conditions and enhances resiliency during adversarial attacks.
\end{theorem}

% \begin{proof}
\textbf{proof}: Under the attacked dynamics \eqref{eq:affine_under_attack},
\[
\dot V_j = L_{f_j}V_j + L_{g_j}V_j\,(u_j+\delta_j(t)).
\]
Using the AR-CLF constraint \eqref{eq:resilient_qp} and the Cauchy--Schwarz inequality, $L_{g_j}V_j\,\delta_j \le \|L_{g_j}V_j\|\,\|\delta_j(t)\|
\le \|L_{g_j}V_j\|\,\gamma_j e^{\kappa_j t}$, we obtain the bound
\begin{equation}
\label{eq:Vdot_compact_final}
\dot V_j \le -\beta_j V_j
-\frac{\|L_{g_j}V_j\|^{2}}{\|L_{g_j}V_j\|+e^{-\alpha t}}\,e^{\rho_j(t)}
+\|L_{g_j}V_j\|\,\gamma_j e^{\kappa_j t}.
\end{equation}
Consider the region $\|x_j\|\ge\varepsilon_j$.  
Assumption~\ref{ass:nonvanishing_clf_gradient} gives $\|L_{g_j}V_j\|\ge c_j$, and since the map 
$s\mapsto s^{2}/(s+e^{-\alpha t})$ is increasing in $s\ge0$, $\frac{\|L_{g_j}V_j\|^{2}}{\|L_{g_j}V_j\|+e^{-\alpha t}}
\ge \bar c_j = \frac{c_j^{2}}{c_j+1}$. Thus, when $\|x_j\|\ge\varepsilon_j$,
\begin{equation}
\label{eq:Vdot_intermediate}
\dot V_j \le -\beta_jV_j - \bar c_j e^{\rho_j(t)} + c_j\gamma_j e^{\kappa_j t}.
\end{equation}
Next, the adaptive law \eqref{eq:rho_dynamics} satisfies $\dot\rho_j(t)=q_j\|L_{g_j}V_j\|\ge q_j c_j
\;\text{whenever } \|x_j\|\ge\varepsilon_j$, which implies the linear growth bound $\rho_j(t)\ge \rho_j(t_0) + q_j c_j (t-t_0)$, over any time interval on which $\|x_j\|\ge\varepsilon_j$.
Since $q_j c_j > \kappa_j$ by design, $e^{\rho_j(t)}$ eventually dominates $e^{\kappa_j t}$.  
Hence, there exists $t_1>0$ such that
\begin{equation}
\label{eq:exp_domination_final}
\bar c_j e^{\rho_j(t)} \ge c_j\gamma_j e^{\kappa_j t},
\qquad \forall\, t\ge t_1 \text{ and }\|x_j\|\ge\varepsilon_j.
\end{equation}
Substituting \eqref{eq:exp_domination_final} into \eqref{eq:Vdot_intermediate} yields $\dot V_j \le -\beta_j V_j,
\forall\, t\ge t_1,\ \|x_j\|\ge\varepsilon_j$. Thus, after a finite time $t_1$, $V_j$ decreases strictly whenever the state lies outside the compact set 
$\mathcal{B}_{\varepsilon_j}=\{x_j:\|x_j\|\le\varepsilon_j\}$.  
Inside $\mathcal{B}_{\varepsilon_j}$, boundedness follows from continuity of the closed-loop dynamics.  
By standard Lyapunov and LaSalle invariance principle, the subsystem trajectories are UUB 
and converge to an invariant set contained in $\mathcal{B}_{\varepsilon_j}$. 
\hfill $\blacksquare$
\section{Simulation and Validation}
A single-bus DC microgrid with two DER-interfacing converters supplying a nonlinear load is used to validate the proposed AR-CLF based QP controller. The switched-circuit model, shown in Fig.~\ref{DC MG_structure}, is simulated in Matlab-Simulink, and all line, filter, and converter parameters are listed in Table~\ref{tab:mg_params}. The control objective is to (i) regulate the DER terminal voltages $v_j$ and the bus voltage $v_b$ to their optimal steady-state values, (ii) ensure bounded converter currents $i_{s_j}$ and load filter current $i_f$, and (iii) maintain closed-loop stability under a wide range of FDI attacks. The steady-state operating point used in all simulations is 
$(x^{*},u^{*})$, given by 
$v_1^{*}=v_2^{*}\approx 24.16~\mathrm{V}$, 
$i_{t_1}^{*}\approx 8.75~\mathrm{A}$, 
$i_{t_2}^{*}\approx 9.25~\mathrm{A}$, 
$v_b^{*}=24~\mathrm{V}$, 
$i_f^{*}=12~\mathrm{A}$, 
$v_l^{*}=12~\mathrm{V}$, 
and $d^{*}=0.5$. The proposed AR-CLF controller is compared against the nominal optimal controller (no resiliency) under different scenarios.
\subsubsection{Case I: Nominal PH-Based Control Under Bounded and Polynomial Attacks}
FDI disturbances are injected at $t=10$\,s and applied directly to the control input channel.

\begin{figure}[t]
    \centering
    \begin{subfigure}{0.49\columnwidth}
        \centering
        \includegraphics[width=\linewidth]{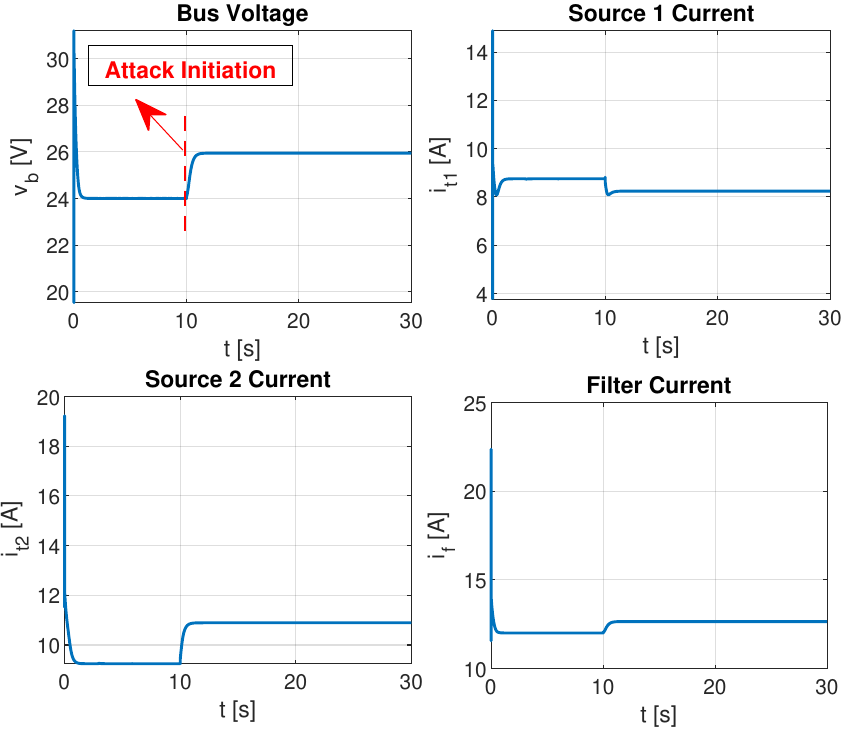}
        \caption{Constant attack}
        \label{fig:conv_constant}
    \end{subfigure}
    \hfill
    \begin{subfigure}{0.49\columnwidth}
        \centering
        \includegraphics[width=\linewidth]{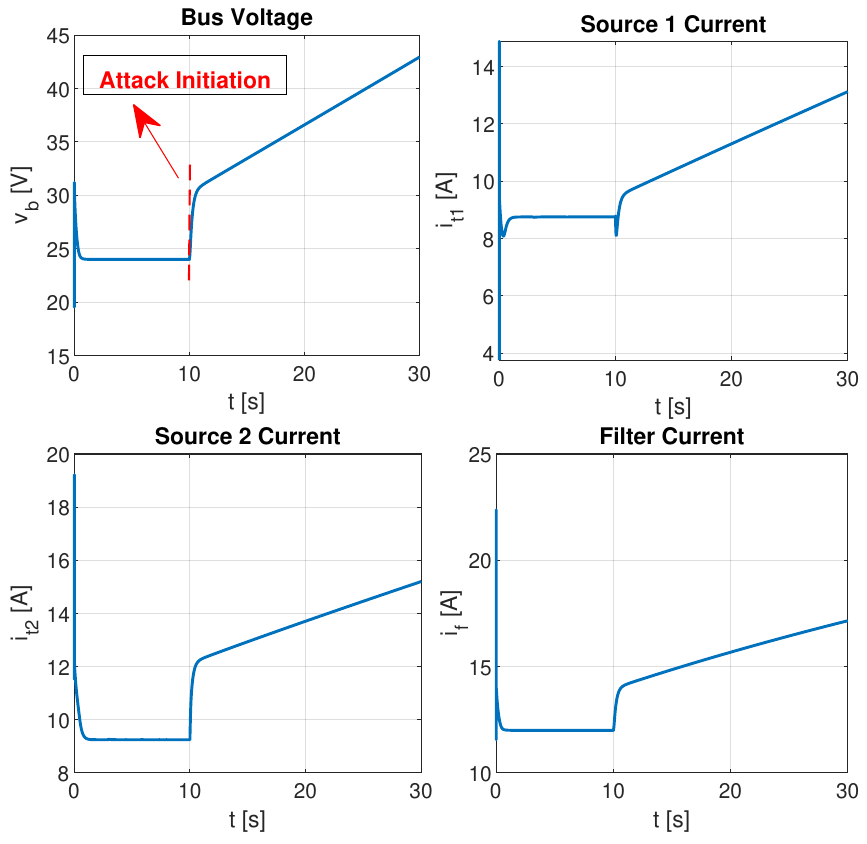}
        \caption{Polynomial attack}
        \label{fig:conv_poly}
    \end{subfigure}

    \caption{Performance of the nominal CLF controller under bounded and polynomially unbounded FDI attacks.}
    \label{fig:conv_attacks}
\end{figure}

I. Bounded attack: A constant bias $\delta(t)=[0.5;0.5;0.5]$ is introduced. As seen in Fig.~\ref{fig:conv_constant}, the nominal controller immediately develops voltage drift and current deviation once the attack begins.

II. Polynomially unbounded attack:  Fig.~\ref{fig:conv_poly} shows that a growing perturbation 
$\delta(t)=1+5[0.2t;0.15t;0.1t]$ quickly drives the system unstable after $t=10$\,s, causing bus-voltage collapse and diverging converter currents. The nominal PH-based controller is reliable only under benign conditions and fails under bounded and growing FDI attacks, motivating attack-resilient control.
\begin{table}[t]
\centering
\caption{Physical parameters of the DC microgrid used in simulation.}
\vspace{-3mm}
\begin{tabular}{c|c|c|c}
\hline
Component & $C$ [mF] & $L$ [mH] & $R$ \\ \hline
DER 1     & 0.49 & 0.09 & $18.78\,\mathrm{m}\Omega$ \\
DER 2     & 0.57 & 0.08 & $17.78\,\mathrm{m}\Omega$ \\
Filter    & 0.47 & 0.16 & $r_l=1\,\Omega$ \\
Bus/Load  & 0.47 & --   & $R_l=2\,\Omega$ \\ \hline
\end{tabular}
\label{tab:mg_params}
\end{table}
% \begin{table}[t]
% \centering
% \caption{DC microgrid parameters and optimal steady-state values used in simulation.}
% \begin{tabular}{c|ccc}
% \multicolumn{4}{c}{\textbf{Component parameters}} \\ \hline
%  & Capacitance [mF] & Inductance [mH] & Resistance \\ \hline
% DER 1 ($j=1$) 
%   & $C_1 = 0.49$ 
%   & $L_1 = 0.09$ 
%   & $R_1 = 18.78~\mathrm{m}\Omega$ \\
% DER 2 ($j=2$) 
%   & $C_2 = 0.57$ 
%   & $L_2 = 0.08$ 
%   & $R_2 = 17.78~\mathrm{m}\Omega$ \\
% Load filter 
%   & $C_l = 0.47$ 
%   & $L_f = 0.16$ 
%   & $r_l = 1~\Omega$ \\
% Bus 
%   & $C_b = 0.47$ 
%   & -- 
%   & $R_l = 2~\Omega$ \\ \hline
% \multicolumn{4}{c}{\textbf{Optimal steady-state values (from code)}} \\ \hline
% State      & \multicolumn{2}{c}{Value} & Units \\ \hline
% $v_1^{*} = v_2^{*}$ & \multicolumn{2}{c}{24.16} & V \\
% $i_{t_1}^{*}$       & \multicolumn{2}{c}{8.75}  & A \\
% $i_{t_2}^{*}$       & \multicolumn{2}{c}{9.25}  & A \\
% $v_b^{*}$           & \multicolumn{2}{c}{24.00} & V \\
% $i_f^{*}$           & \multicolumn{2}{c}{12.00} & A \\
% $v_l^{*}$           & \multicolumn{2}{c}{12.00} & V \\
% \end{tabular}
% \label{tab:mg_params_equil}
% \end{table}
% \vspace{-15mm}
\vspace{-2mm}
\subsubsection{Case II: Proposed AR-CLF based QP Control Under Polynomial and Exponential Unbounded Attacks}
\begin{figure}[t]
    \centering
    \begin{subfigure}{0.49\columnwidth}
        \centering
        \includegraphics[width=\linewidth]{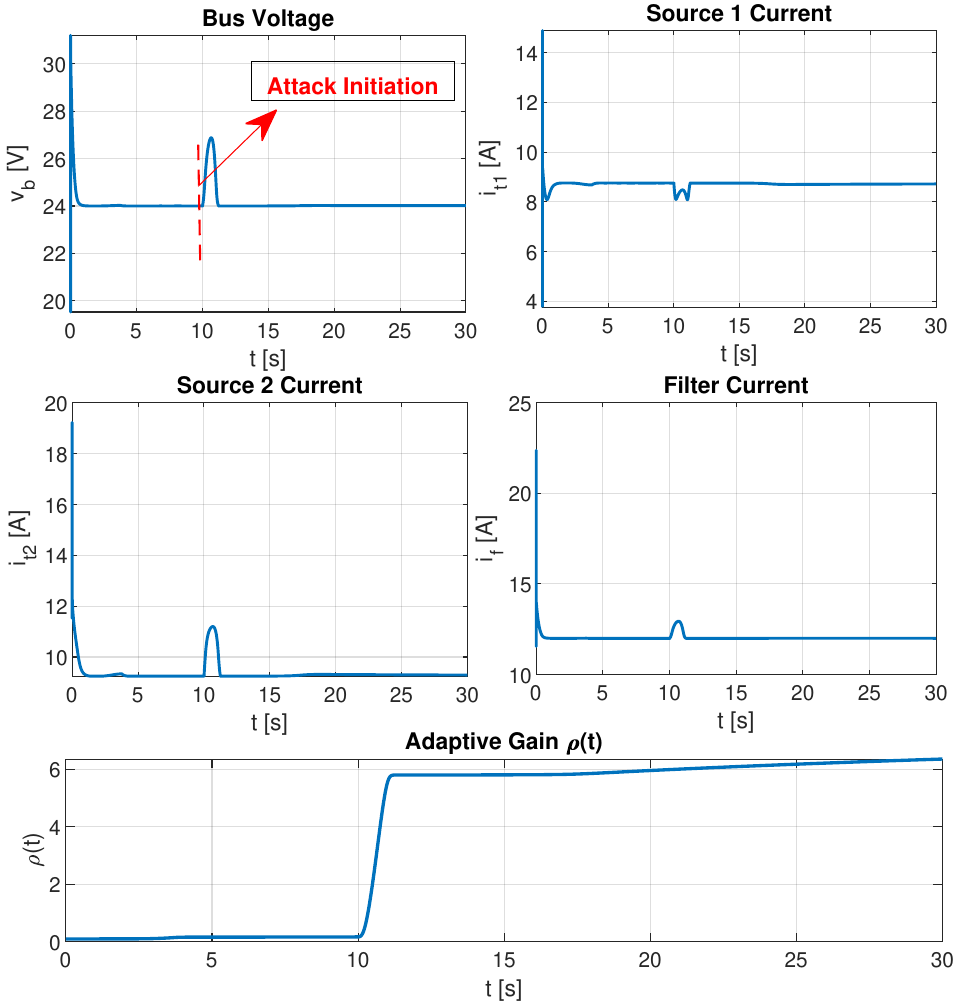}
        \caption{Polynomial attack}
        \label{fig:prop_poly}
    \end{subfigure}
    \hfill
    \begin{subfigure}{0.49\columnwidth}
        \centering
        \includegraphics[width=\linewidth]{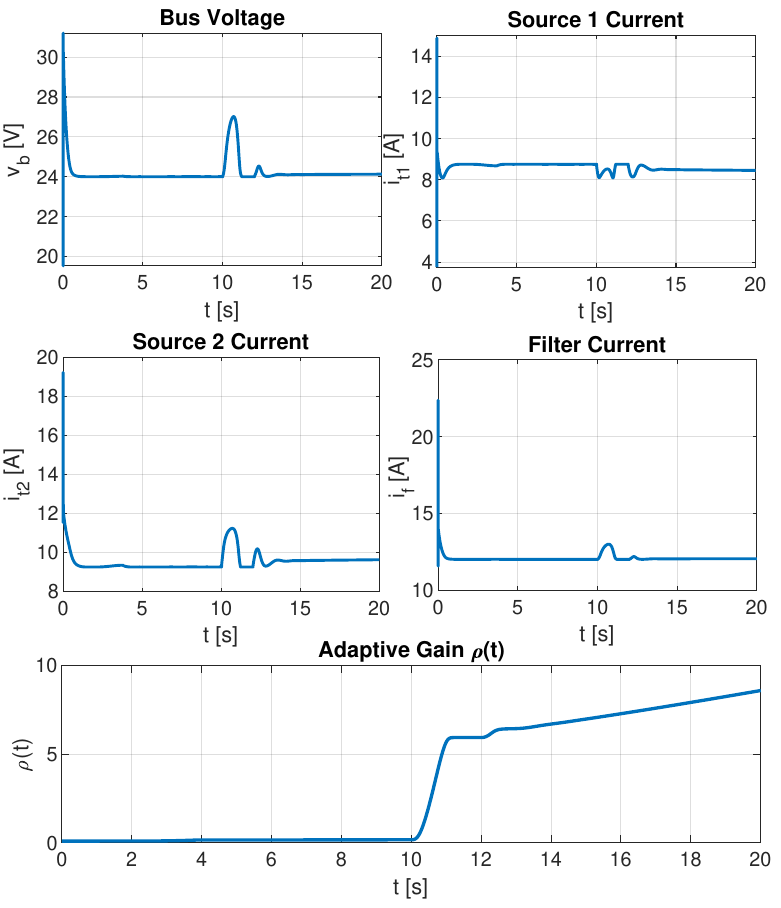}
        \caption{Exponential attack}
        \label{fig:prop_expo}
    \end{subfigure}

    \caption{Performance of the proposed AR-CLF controller under polynomially and exponentially unbounded FDI attacks.}
    \label{fig:proposed_attacks}
\end{figure}

We evaluate the proposed AR-CLF based QP controller under the same attack injection window ($t=10$\,s) using the following scenarios:

I. Polynomially unbounded attack: With 
$\delta(t)=1+5[0.2t;\,0.15t;\,0.1t]$, the proposed controller maintains bounded converter currents and preserves voltage regulation (Fig.~\ref{fig:prop_poly}). Unlike the nominal controller, which diverges under this disturbance, the adaptive gain $\rho(t)$ grows fast enough to suppress the increasing attack magnitude, ensuring convergence toward an attack-inflated invariant set.

% II. Exponentially unbounded attack: Under the severe perturbation  $\delta(t)=0.15[\exp(0.3t);\exp(0.3t);\exp(0.3t)]$, the AR-CLF controller remains stable (Fig.~\ref{fig:prop_expo}). The adaptive law autonomously amplifies $\rho(t)$ as the attack escalates, dominating the adversarial term in the CLF derivative and guaranteeing UUB of all states. Bus voltage, DER terminal voltages, and all currents remain bounded and close to their nominal steady-state values. The results show that AR-CLF ensures decentralized resilience against a broad range of FDI attacks, including exponentially unbounded attacks.

II. Exponentially unbounded and stochastic attacks: We consider heterogeneous FDI attacks of the form
$\delta(t)=
\big[
0.15(2+4e^{0.3(t-t_1)})\mathbb{I}_{t\ge t_1},
0.15(5+5e^{0.4(t-t_2)})\mathbb{I}_{t\ge t_2},
0.10(3+10e^{0.2(t-t_3)})\mathbb{I}_{t\ge t_3}
\big]
+\delta_{\mathrm{noise}}(t)$,
with channel-dependent gains, growth rates, asynchronous initiation times, and $\delta_{\mathrm{noise}}(t)$ represents additive stochastic disturbances (e.g., Gaussian noise).
Under these severe perturbations, the proposed AR-CLF based QP controller remains stable
(Fig.~\ref{fig:prop_expo}). As the attacks escalate, the adaptive law autonomously amplifies the compensation parameter $\rho(t)$, ensuring the adversarial terms are compensated in the CLF dissipation inequality and guaranteeing UUB of all closed-loop states. Consequently, the DC bus voltage, DER terminal voltages, and line currents remain bounded and close to their nominal steady-state values despite asynchronous, stochastic, and exponentially unbounded FDI attacks. The DC bus voltage deviation remains below $12\%$ of nominal, and all currents stay bounded with peaks under $11\%$ during the attack, demonstrating that the proposed controller ensures decentralized resilience against a broad class of realistic cyberattacks beyond bounded or synchronized models.

% \begin{figure}[t]
%     \centering
%     \includegraphics[width=\columnwidth]{Con_Constant.pdf}
    
%     \caption{Conventional-CLF under constant attack.}
%     \label{Conv_Constant}
% \end{figure}

% \begin{figure}[t]
%     \centering
%     \includegraphics[width=\columnwidth]{Conv_polly.pdf}
    
%     \caption{Conventional-CLF under polynomially unbounded attack.}
%     \label{Conv_polly}
% \end{figure}
% \begin{figure}[t]
%     \centering
%     \includegraphics[width=\columnwidth]{Proposed_Constant.pdf}
    
%     \caption{Proposed-CLF under constant attack.}
%     \label{Proposed_Constant}
% \end{figure}
% \begin{figure}[t]
%     \centering
%     \includegraphics[width=\columnwidth]{Proposed_polly.pdf}
    
%     \caption{Proposed-CLF under polynomially unbounded attack.}
%     \label{Proposed_polly}
% \end{figure}

% \begin{figure}[t]
%     \centering
%     \includegraphics[width=\columnwidth]{Proposed_expo.pdf}
    
%     \caption{Proposed-CLF under exponentially unbounded attack.}
%     \label{Proposed_expo}
% \end{figure}
\vspace{-1em}
\section{Conclusion}
This paper presents a fully decentralized AR--CLF--based QP control framework for nonlinear DC microgrids under a broad class of FDI attacks, including polynomially and exponentially unbounded disturbances. Using port--Hamiltonian modeling and an adaptive law, the proposed controller guarantees large-signal stability from local measurements only, without global communication, centralized coordination, or explicit attack detection. Rigorous Lyapunov analysis establishes uniform ultimate boundedness, and simulations demonstrate that, unlike the nominal PH controller, the AR--CLF--QP preserves voltage regulation, bounds currents, and autonomously compensates for escalating attacks, providing a scalable and resilient solution for adversarial next-generation DC microgrids.
\ifCLASSOPTIONcaptionsoff
  \newpage
\fi
\vspace{-1em}
\bibliographystyle{IEEEtran}

\bibliography{References}

\end{document}